\documentclass[11pt,letterpaper]{article}

\pdfoutput=1  

\usepackage{palatino}
\usepackage[left=1in,top=1in,right=1in,bottom=1in]{geometry}
\usepackage{amsmath}
\usepackage{amssymb}
\usepackage{amsbsy}
\usepackage{amsthm}
\usepackage{epsfig}
\usepackage{clrscode}
\usepackage{setspace}
\usepackage{multirow}
\usepackage{url}
\usepackage{paralist}
\usepackage{wrapfig,boxedminipage}
\usepackage{graphicx}
\usepackage{subcaption}
\usepackage{float}
\usepackage{xcolor}
\usepackage{pgfplotstable}
\usepackage{hyperref}
\usepackage{enumitem} 
\usepackage{adjustbox}
\usepackage{booktabs}
\usepackage{csvsimple}
\usepackage{tikz}
\usepackage{dcolumn}
\usepackage{setspace} 
\usepackage{indentfirst}

\usepackage{graphicx}

\usepackage{bibentry}

\usepackage[round]{natbib}

\newcommand{\nc}{\newcommand}

\nc{\bX}{{\bf X}}
\nc{\lhat}[1][i]{\hat\lambda_{#1}^{-1(g)}}
\nc{\what}[1][j]{\hat\omega_{#1}^{-1(g)}}
\nc{\Li}{\hat\Lambda^{-1(g)}}
\nc{\Oi}{\hat\Omega^{-1(g)}}
\nc{\diag}[1]{\text{diag}\left(#1\right)}
\nc{\Siginv}{\Sigma^{-1}}
\nc{\Ominv}{\Omega^{-1}}
\nc{\bone}{{\bf 1}}

\newcommand{\beq}{\begin{equation}}
\newcommand{\eeq}{\end{equation}}

\newcommand{\ben}{\begin{enumerate}}
\newcommand{\een}{\end{enumerate}}

\providecommand{\keywords}[1]{\textbf{\textit{Keywords and phrases: }} #1}

\begin{document}

\title{Monotone function estimation in the presence of extreme data coarsening: Analysis of preeclampsia and birth weight in urban Uganda}
\author{Jennifer E.~Starling\footnote{Corresponding author: jstarling@utexas.edu}, Catherine E. Aiken, Jared S. Murray, \\ Annettee Nakimuli, and James G.~Scott}

\maketitle

\begin{abstract}


This paper proposes a Bayesian hierarchical model to characterize the relationship between birth weight and maternal pre-eclampsia across gestation at a large maternity hospital in urban Uganda.  Key scientific questions we investigate include: 1) how pre-eclampsia compares to other maternal-fetal covariates as a predictor of birth weight; and 2) whether the impact of pre-eclampsia on birthweight varies across gestation.  Our model addresses several key statistical challenges: it correctly encodes the prior medical knowledge that birth weight should vary smoothly and monotonically with gestational age, yet it also avoids assumptions about functional form along with assumptions about how birth weight varies with other covariates.  Our model also accounts for the fact that a high proportion (83\%) of birth weights in our data set are rounded to the nearest 100 grams.  Such extreme data coarsening is rare in maternity hospitals in high resource obstetrics settings but common for data sets collected in low and middle-income countries (LMICs); this introduces a substantial extra layer of uncertainty into the problem and is a major reason why we adopt a Bayesian approach.

Our proposed non-parametric regression model, which we call Projective Smooth BART (psBART), builds upon the highly successful Bayesian Additive Regression Tree (BART) framework.  This model captures complex nonlinear relationships and interactions, induces smoothness and monotonicity in a single target covariate, and provides a full posterior for uncertainty quantification.  The results of our analysis show that pre-eclampsia is a dominant predictor of birth weight in this urban Ugandan setting, and therefore an important risk factor for perinatal mortality.




\end{abstract}


\newpage

\onehalfspace

\begin{centering}
\section{Introduction} \end{centering}


An ongoing research and policy challenge in global public health is to address high rates of mortality among babies born at low birth weights in lower and middle income countries (LMICs).  Neonatal survival depends on a range of factors, with birth weight as a key determinant \citep{lawn2014}; babies born at low birth weights for their gestational age have markedly higher risk of perinatal death \citep{katz2013, reinebrant2018, kozuki2017}.  Thus elucidating risk factors for low birth weight in different populations globally is a key research goal for understanding and ultimately preventing neonatal mortality.

While some risk factors for low birth weight are common across human populations (e.g.~sex and gestational age at delivery), other risk factors are highly context-dependent \citep{fried2018, dasilvalopes2017}.  In this paper we focus specifically on data from sub-Saharan Africa, where maternal pre-eclampsia---a disorder of pregnancy characterized by the onset of high blood pressure and proteinuria---is common, severe, under-recognized, and often untreated \citep{nakimuli2014,nakimuli2016,firoz2011}, and is a known risk factor for low birth weight \citep{odegard2000}.  In urban Uganda, the perinatal death rate in pregnancies affected by pre-eclampsia in urban Uganda is twice that in normotensive women \citep{nakimuli2015}, with some evidence suggesting a perinatal death rate of over 20\% in pregnancies complicated by maternal pre-eclampsia \citep{kiondo2014}.  

Statistically, we can think of expected birth weight ($y$, measured in kg) as a regression function $E(y) = f(t,x)$, where $t$ is gestational age at delivery and $x$ is a vector of maternal-fetal characteristics, including presence of maternal pre-eclampsia.  The fundamental scientific problem we address in this paper is to provide provide better estimates of $f(t,x)$, thereby allowing us to compare the influence on birthweight of pre-eclampsia versus other maternal-fetal risk factors in an urban Ugandan population where these risk factors are not well characterized.  We propose a model for doing so, and we use this model to analyze data from a very recent study of 2,444 pregnancies from Mulago Hospital, the largest public hospital in Uganda.  The results of our analysis show that maternal pre-eclampsia is the dominant predictor of having low birth weight for gestational age in this population and therefore an important risk factor for perinatal mortality.   We also show that in this setting, the influence of pre-eclampsia on birth weight is higher at lower gestational ages.

\begin{figure}[t]
		\centering
			\includegraphics[width=0.5\textwidth]{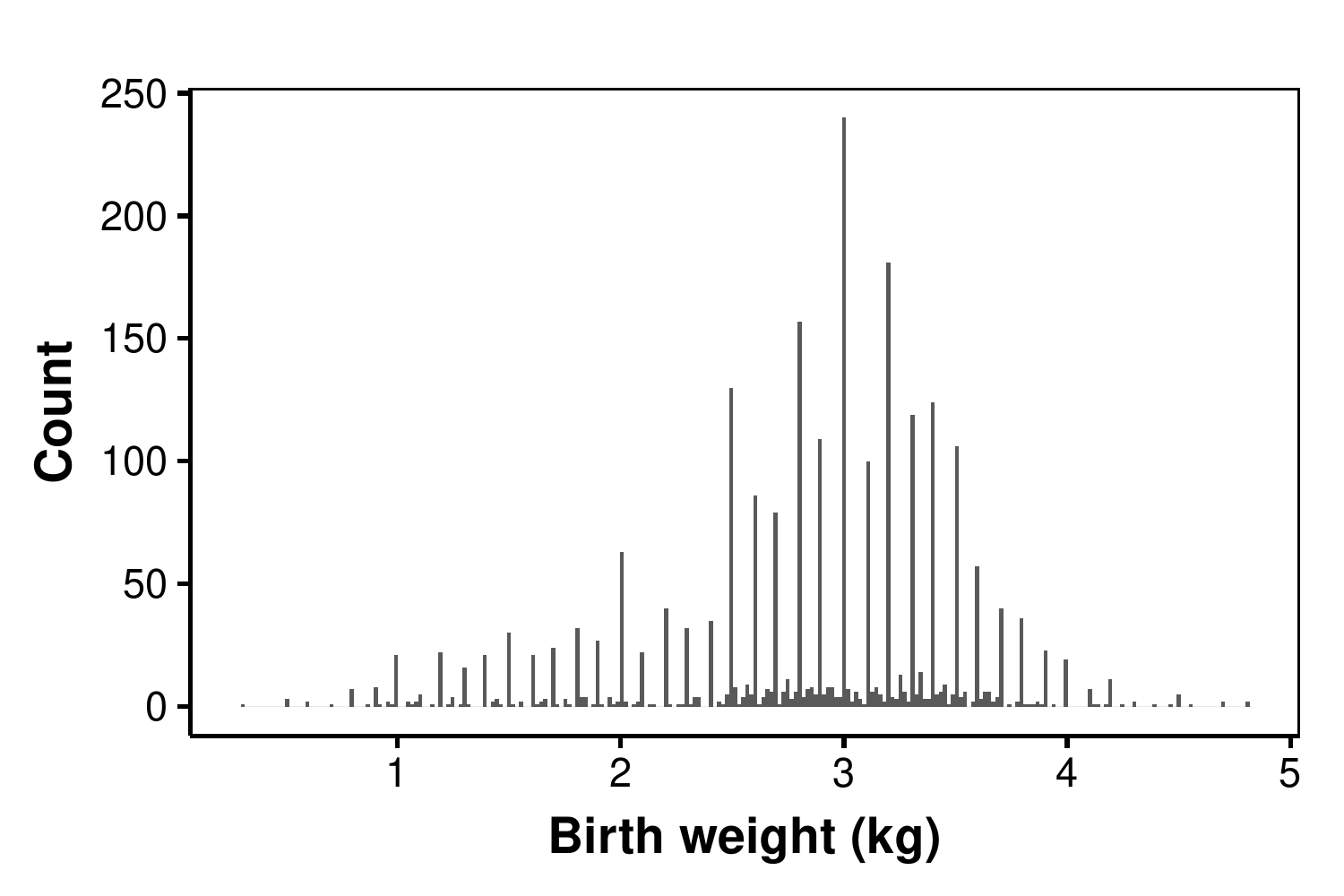}
		\caption{Illustrates heaping in the birth weight outcome.  Of the 2,444 observations, 2,033 (83\%) are observed at birth weights that are even to the nearest 100 grams and so potentially rounded.}
		\label{fig:mulag-bin-hist}
	\end{figure}

To realize these scientific goals, our analysis was forced to reckon with four statistical challenges, which our proposed model has been designed specifically to address.
\begin{enumerate}
\item Prior scientific knowledge dictates that expected birth weight should increase smoothly with gestational age, implying that $f(t,x)$ should be restricted to the space of monotone increasing functions in $t$.
\item In many obstetric data sets collected in LMICs, birth weights are often recorded to the nearest 100 grams, resulting in heaped data with a high proportion of data coarsening---in our case, 83\% of observations. (See Figure \ref{fig:mulag-bin-hist} on page \pageref{fig:mulag-bin-hist}.)  
\item Our model should make no strong assumptions about how $f(t,x)$ depends on other maternal-fetal characteristics $x$, since these relationships are not well understood---especially in LMICs.
\item Our model should produce valid uncertainty estimates, even in the presence of data coarsening.
\end{enumerate}

As we will argue, no existing method in the literature can successfully meet all four challenges.   Owing to this fact, and also owing to the lack of good data, current medical literature does not provide an especially nuanced characterization of $f(t,x)$ in LMICs.   Much of the previous work that investigates risk factors for low birth weight in LMICs either excludes pre-eclamptic cases, fails to consider pre-eclampsia as a factor, or lacks data entirely on pre-eclampsia \citep{he2018,muhihi2016,mekonen2015}.  Recent work by \citet{nakimuli2019} seeks to delineate the influence of pre-eclampsia on birth weight in urban Uganda across viable gestation, using prospectively collected data specifically focused on pregnancies impacted by pre-eclampsia.  Their approach relies on a multi-step model selection process, manual specification of which interactions to include in the model, and stepwise selection of basis functions.  It also fails to account correctly for monotonicity or data coarsening.

To address these challenges, we leverage the BART with Targeted Smoothing (tsBART) framework of \citet{starling2019}, which is itself based on the Bayesian Additive Regression Trees (BART) model of \citet{chipman2010}.  TsBART, like BART, is a ``sum of weak learners'' model; it differs from BART in allowing $y$ to vary smoothly in a specific target covariate $t$, but not necessarily in other covariates $x$.  This is ideally suited to our application, where expected birth weight $f(t,x)$ should vary smoothly with gestational age $t$.  The tsBART model constructs $f(t,x)$ by summing a large number of binary regression trees, each of which is encouraged by a strong prior to have relatively few splits.  The trees split on $x$ only, while terminal tree nodes are parametized by Gaussian processes in $t$, resulting in a prior for $f(t,x)$ which is smooth in $t$ but not necessarily in $x$.  Modifying BART to incorporate this form of targeted smoothing can usefully encode prior knowledge about smoothness, aid interpretability for clinicians and patients, and provide gains in mean-squared error \citet{starling2019}.

To analyze pre-eclampsia's impact on birth weight in our data set, we must further extend tsBART in two important ways.  First, we use the projective Gaussian process approach of \citet{lin2014} to introduce an additional monotonicity constraint in $t$, while not enforcing monotonicity or even smoothness over other covariates $x$.  This reflects the additional prior knowledge that average birth weights should increase with gestational age.  Correctly incorporating this structural assumption is especially helpful for sharpening estimates of the pre-eclampsia effect on birth weight since gestational age at delivery and the presence of pre-eclampsia are correlated.  Second, we account for the coarseness of birthweight data by introducing latent variables representing the true unknown weights, and we sample from the posterior distribution over these weights, given the rounded-off observations.  This correctly incorporates the additional layer of uncertainty that arises from rounded-off birth weights.

The rest of the paper proceeds as follows.  Section \ref{sec-apps} provides an overview of the dataset and the pre-eclampsia modeling problem.  Section \ref{sec-meths} details the model, placing it into the context of three relevant lines of prior work in the statistical literature: BART, monotone function estimation, and data coarsening.  Section \ref{sec-sims} presents an illustration of the importance of accounting for coarseness, and the results of a simulation study showing advantages of incorporating monotonicity constraints when appropriate.  Section \ref{sec-results} presents our core scientific contribution: our analysis of pre-eclampsia and birth weight for the Mulago hospital data.  Section \ref{sec-discussion} concludes with discussion.  An R package implementing our methods is described in Section \ref{sec-supp}.

\begin{centering}\section{Pre-eclampsia in urban Uganda: background and data} \label{sec-apps} \end{centering}


\begin{table}[p]
    \caption{ \label{tbl:cohort}
	Cohort characteristics for the pre-eclampsia dataset.  PE indicates the presence of pre-eclampsia.  Numbers in parentheses are percentages with respect to the given cohort.}
\centering
 \footnotesize
\begin{tabular}{lrrrr}
 Characteristic & (N = 2,444) & No PE (N = 1,456) & PE (N = 988) & P-value \\ 
  \toprule
Gestational age in weeks, n (\%) &  &  &  & $<$ 0.0001 \\ 
   \hline
28-31 & 117 (4.79) & 16 (1.10) & 101 (10.22) &  \\ 
  32-37 & 555 (22.71) & 142 (9.75) & 413 (41.80) &  \\ 
  38-39 & 892 (36.50) & 638 (43.82) & 254 (25.71) &  \\ 
  40-42 & 823 (33.67) & 620 (42.58) & 203 (20.55) &  \\ 
  43 & 57 (2.33) & 40 (2.75) & 17 (1.72) &  \\ 
   \hline
Maternal age in years, n (\%) &  &  &  & $<$ 0.0001 \\ 
   \hline
(13,20] & 840 (34.37) & 640 (43.96) & 200 (20.24) &  \\ 
  (20,30] & 1,335 (54.62) & 768 (52.75) & 567 (57.39) &  \\ 
  (30,40] & 259 (10.60) & 47 (3.23) & 212 (21.46) &  \\ 
  (40,46] & 10 (0.41) & 1 (0.07) & 9 (0.91) &  \\ 
   \hline
Maternal job type, n (\%) &  &  &  & $<$ 0.0001 \\ 
   \hline
Skilled & 376 (15.38) & 206 (14.15) & 170 (17.21) &  \\ 
  Unskilled & 1,214 (49.67) & 780 (53.57) & 434 (43.93) &  \\ 
  Unemployed & 854 (34.94) & 470 (32.28) & 384 (38.87) &  \\ 
   \hline
Infant sex, n (\%) &  &  &  & = 1.0000 \\ 
   \hline
Female & 1,240 (50.74) & 729 (50.07) & 511 (51.72) &  \\ 
  Male & 1,204 (49.26) & 727 (49.93) & 477 (48.28) &  \\ 
   \hline
Ganda ethnicity, n (\%) &  &  &  & = 0.0238 \\ 
   \hline
No & 1,023 (41.86) & 637 (43.75) & 386 (39.07) &  \\ 
  Yes & 1,421 (58.14) & 819 (56.25) & 602 (60.93) &  \\ 
   \hline
Maternal HIV, n (\%) &  &  &  & = 0.7897 \\ 
   \hline
Not Positive & 2,313 (94.64) & 1,376 (94.51) & 937 (94.84) &  \\ 
  Positive & 131 (5.36) & 80 (5.49) & 51 (5.16) &  \\ 
   \hline
Max Systolic BP Delivery BP &  &  &  & $<$ 0.0001 \\ 
   \hline
mean $\pm$ sd & 137.75 $\pm$ 33.39 & 113.67 $\pm$ 9.93 & 173.24 $\pm$ 22.29 &  \\ 
   \hline
Max Diastolic Delivery BP &  &  &  & $<$ 0.0001 \\ 
   \hline
mean $\pm$ sd & 88.68 $\pm$ 25.50 & 70.10 $\pm$ 8.06 & 116.07 $\pm$ 15.90 &  \\ 
   \hline
Max Proteinuria, n (\%) &  &  &  & $<$ 0.0001 \\ 
   \hline
0 & 1,425 (58.31) & 1,425 (97.87) & 0 (0.00) &  \\ 
  1 & 29 (1.19) & 29 (1.99) & 0 (0.00) &  \\ 
  2 & 348 (14.24) & 0 (0.00) & 348 (35.22) &  \\ 
  3 & 366 (14.98) & 0 (0.00) & 366 (37.04) &  \\ 
  4 & 276 (11.29) & 2 (0.14) & 274 (27.73) &  \\ 
   \hline
Parity, n (\%) &  &  &  & $<$ 0.0001 \\ 
   \hline
Multiparous & 554 (22.67) & 53 (3.64) & 501 (50.71) &  \\ 
  Primiparous & 1,890 (77.33) & 1,403 (96.36) & 487 (49.29) &  \\ 
   \bottomrule
\end{tabular}
 \smallskip
\end{table}

Pre-eclampsia is diagnosed by the presence of hypertension during pregnancy, in combination with proteinuria or seizure activity \citep{acog2013}.  The initiating event in pre-eclampsia is incomplete conversion of spiral arteries during early pregnancy due to insufficient invasion of vessel walls by the extra-villous trophoblast \citep{nakimuli2019,moffett2015}.  Poor materno-fetal vascular connections in the placenta result in restricted growth and insufficient transfer of nutrients to the growing fetus \citep{burton2017}.  Severe pre-eclampsia is prevalent in sub-Saharan African women, possibly due to a combination of genetic factors, lack of access to quality antenatal care, and poor baseline health status \citep{nakimuli2016}.

Much of the previous work that looks at factors influencing birth weight in LMICs do not include pre-eclampsia, often due to unavailability of data \citep{he2018,muhihi2016,mekonen2015}.  Pre-eclampsia is under-diagnosed and under-treated in many LMICs, complicating efforts to study it.  One exception is the study by \citet{nakimuli2019}, who prospectively collected a sample that entailed careful phenotyping of preeclampsia cases alongside a control set of normotensive pregnancies.  We perform our analysis using this dataset.  The original analysis in \citet{nakimuli2019} relies on a multi-step model selection process, as well as several model choices related to basis function selection and manual specification of interactions.  Moreover, it does not address monotonicity constraints or heaped observations.  The analysis of this data set can therefore benefit greatly from our new statistical approach, specifically designed to address these features in the birth weight outcome.

The data set from \citet{nakimuli2019} was collected at the Mulago National Referral and Teaching Hospital in Kampala, Uganda.  Mulago Hospital is a tertiary referral center for Uganda, and is one of the largest obstetrics centers globally, accommodating approximately 30,000 deliveries a year.   Women were recruited into the study in three waves: during August 2010 -- June 2011, September 2014 -- December 2016, and July 2019.    Outcomes, delivery procedures, and standards of care received by patients were similar across waves.  Data were collected by research midwives at the time of initial presentation using a researcher-administered questionnaire, with further information gathered from each patient's medical records.  Each of the 2,444 observations in the dataset is associated with a singleton pregnancy.  Our analysis also includes stillborn babies judged by the clinician to have died shortly before delivery.  We also include cases of early neonatal death.  As our focus is specifically on pre-eclampsia rather than other hypertensive disorders, patients with known chronic hypertension or renal disease were excluded from the dataset.

The response was birth weight, measured in kilograms.  Figure \ref{fig:mulag-bin-hist} shows the strong presence of heaping in this dataset.  Of the 2,444 observations, 2,033 (83\%) have recorded birth weights that are even to the nearest 0.1 kg (100 grams) and so potentially rounded.  

Each observation contains the gestational age at time of delivery in weeks, calculated for each pregnancy based on last menstrual period or ultrasound scan.  We consider only singleton infants born between 28 and 42 weeks estimated gestation.  Infant sex was designated as male or female at delivery.

Pregnancies were classified by presence or absence of pre-eclampsia. The sample consists of 988 pregnancies with pre-eclampsia and 1,456 without.  On recruitment into the study, women were designated as affected by pre-eclampsia or not based on clinical notes.  Diagnostic criteria were based on a context-appropriate adaptation of the ACOG Task Force Report on Hypertension in Pregnancy \citep{acog2013}.  Pre-eclampsia was diagnosed where systolic blood pressure had been measured as $\geq$ 140, or diastolic as $\geq$ 90, on at least two occasions four hours apart, in conjunction with either $\ge$1+ proteinuria on dipstick or clinical seizure activity.  As in other low-resource settings \citep{ukah2017}, routine blood tests are not performed on all women presenting with hypertension and proteinuria in the study center, so we were unable to apply ACOG criteria for diagnosis of pre-eclampsia which relies on laboratory tests.  


Maternal-fetal characteristics were derived directly from clinical notes or information provided from patients (Table \ref{tbl:cohort}).  Maternal covariates include maternal age in years, parity, whether the mother belonged to the predominant local Ganda ethnicity, and maternal HIV status.  Maternal occupation was self-reported by patients, classified using ISCO-08 classification \citep{zeitlin2016}, and grouped into three categories: professional, skilled, and unskilled/no occupation.  Whether women had experienced severe febrile illness during pregnancy was recorded as a proxy to represent malaria during pregnancy \citep{odongo2016}; we exclude this covariate as it was only gathered for patients in wave 3, and previous work  \citep{nakimuli2019} found that presence of febrile illness during pregnancy was not a significant predictor of birth weight.

\bigskip

\begin{centering}
\section{Methods} \label{sec-meths} 
\end{centering}


Our model, Projective Smooth BART (psBART), builds on BART with Targeted Smoothing (tsBART) \citep{starling2019} and the projective Gaussian process approach of \citet{lin2014}.  The tsBART method is a Bayesian regression model which extends the Bayesian Additive Regression Tree framework introduced by \citet{chipman2010} to induce smoothness in a single 'target' covariate.  We review relevant literature and the tsBART model, then introduce the Projective Smooth BART model.  Our methods are implemented in the \textbf{tsbart} R package. \\

\subsection{Connection with existing work} \label{ss:exwork}


\textbf{Monotone function estimation.} There is a rich body of literature on monotone function estimation in both frequentist and Bayesian frameworks and we give only a brief overview here.  A common approach in the frequentist setting is to minimize a least-squares objective function subject to constraints \citep{barlow1972,robertson1998}.  Recent literature in this area focuses on univariate function estimation, often in the non-parametric dose-response curve setting \citep{bhattacharya2010, bhattacharya2011, bhattacharya2013}.  \citet{mammen1991} used of a smoothing step via application of a kernel estimator.  \citet{cherno2009} showed that applying monotonicity to some unrestricted initial estimator, such as kernels or splines, results in improved estimation error. \citet{ramsay1988} suggest integrated splines, where monotonicity is enforced via basis functions with non-negative coefficients.  

Spline-based approaches have been employed in both frequentist and Bayesian settings. \citet{shively2009} suggests Bayesian spline regression with a mixture of constrained Gaussians as the prior.   \citet{bornkamp2009} model functions as a mix of shifted and scaled parametric probability distribution functions with a random probability measure as the prior for mixing.  \citet{neelon2004} build a piecewise linear model with built-in monotonicity constraint of non-negative slopes, with large number of fixed cutpoints, and an autoregressive smoothing parameter for the slope values.

These models described so far focus on the univariate setting.  There is also recent work on monotone multiple regression.  \citet{saarela2011} define a monotone piecewise constant regression surface using random point locations and associated function levels for monotone multiple regression. \citet{bornkamp2010} enforce monotonicity by considering linear combinations of monotone components with non-negative coefficients; this work does lose some generality, as summing monotone main effects and interactions doesn't define a general monotonic relationship.  \citet{shively2011} uses a Bayesian approach to multiple regression with free-knot and fixed-knot regression splines.  All of these approaches consider monotonicity over all covariates included in the model.

Our work is most closely aligned with the projective Gaussian process approach of \citet{lin2014}, who assume a Gaussian process prior $f \sim GP(\mu,R)$.  The posterior distribution of $f$, another Gaussian process, is then projected onto the space of monotone functions.  We will explore this connection in Section \ref{sec-meth-b}.

\textbf{Data coarsening.} For handling the data coarsening from birth weight rounding, we build on definitions, terminology, and assumptions from \citet{heitjan1991}.  Per \citet{heitjan1991}, data coarsening at random is a version of missingness at random.  We draw on the classic Bayesian missing data literature \citep{rubin1976, rubin1987}, which defines a general model for missing data where randomness due to missing values is explicitly incorporated into the model.

\textbf{Bayesian Additive Regression Trees.} Our paper adds to a body of research on the Bayesian tree-modeling framework.  Smooth extensions of BART have also been proposed by \citet{linero2017} who smooth a regression tree ensemble by randomizing the decision rules at internal nodes of the tree.  This framework induces smoothness over all covariates instead of a single one, by replacing the step function induced by the binary regression trees with sigmoids.  

BART has proven successful in a range of applications, including prediction and classification \citep{chipman2010, murray2017log, linero2017, linero2018, hernandez2018}, survival analysis \citep{sparapani2016}, and causal inference \citep{hill2011, hahn2017, logan2017, sivaganesan2017}. There is ongoing work developing accelerated BART methods for fast posterior estimation \citep{he2019}.  BART has been modified for monotone function estimation \citep{chipman2016b}, though this implementation is monotone in all covariates and doesn't handle data coarsening, making it inappropriate for our setting.

None of these approaches meets all of our needs in this application.  We require a method that is fully nonparametric in all covariates, as relationships between elements of the maternal-fetal dyad and interactions between those covariates and gestation are not well-understood a priori.  Spline-based methods induce the desired smoothness, but require choice of basis functions and selection of knots or additional model complexity for free knots, leading to a potentially very large combinatorial search over possible model specifications.  Moreover, expected birth weight should vary smoothly and monotonically over gestation without imposing smoothness or monotonicity over the other features of the maternal-fetal dyad.  Traditional approaches based on step functions do not guarantee smoothness. Finally, our model must be capable of handling the coarsening in our data set that we have already described, while still producing valid uncertainty estimates. 

\subsection{Overview of tsBART} \label{sec-meth-a} 

Our approach buildings upon the tsBART model, which we now briefly summarize.  The tsBART prior \citep{starling2019} is an extension of BART \citep{chipman2010} for estimating non-linear functions that are smooth in a single target covariate $t$.  It was specifically designed for the scenario common in obstetrics applications, where outcomes vary smoothly in gestational age but not necessarily other covariates; this makes it a natural starting point for our analysis of the Mulago data.

Consider a scalar response, where $y_i = f(t_i,x_i) + \epsilon_i$, where $x_i$ is a vector of covariates and $t_i$ is over a discrete mesh $t \in {t_1,\hdots,t_T}$.  In this model, the scalar node-level parameters $\mu_{hj}$ are replaced with univariate functions $\mu_{hj}(t)$ in the terminal nodes.  Formally, let each observation $i$ consist of $(t_i, x_i, y_i)$ for $i = 1, \ldots, N$.  Let
	\begin{align}
		y_{i} &= f(t_i,x_i) + \epsilon_{i}, \quad \epsilon_{i} \stackrel{iid}{\sim} \text{N}(0, \sigma^2)  \\
		f(t_i, x_i) &= \sum_{j=1}^{m} g(t_i, x_i; T_j, M_j) \nonumber \,
	\end{align}
where $T_j$ are again binary trees.  The tsBART prior is identical to the BART prior for tree space and error variance, while independent Gaussian leaf priors are replaced by a collection of Gaussian processes in $t$: $M_j = \{\mu_{1j}(t), \ldots, \mu_{b_jj}(t) \}$, with each function $\mu_{hj}(t)$ associated with one terminal node.
$$
\mu(t) \sim \text{GP}\left(0, C_\theta(t,t') \right) \, ,
$$
where $C_\theta(t, t')$ is the squared exponential covariance kernel with hyperparameter $\theta$, which can be either chosen based on prior knowledge or tuned using the data.  We refer interested readers to \citet{starling2019} for a discussion of hyperparameter selection, and to \citet{chipman2010} for detail on the original BART model.

\subsection{Latent variable model for coarsened data} \label{sec-meth-c} 

Now consider the heaped data setting where observations are coarse, in the terminology of \citet{heitjan1991}, with some observations $y$ rounded to a common precision $c$.  There are some observations that are clearly not rounded; however, for observations evenly divisible by $c$, it is unknown whether they are coarse or precisely measured.  We assume coarseness at random, as in \citet{heitjan1991}.

Formally, let $t_i$ be observed gestational age and $x_i$ a vector of other maternal-fetal covariates.   Let $\tilde{y}_i$ denote the observed values of birth weight, where some $\tilde{y}_i$ may be rounded to the nearest $c$.  Then define
\begin{align*}
	\gamma_i = \begin{cases}
		1 \text{, if } \tilde{y}_i = \text{round}(\tilde{y}_i,c) \\
		0 \text{, otherwise,}
	\end{cases}
\end{align*}
so that our birthweight data consists of the $(y_i, \gamma_i)$ pairs.  We now let $y_i$ represent the (possibly unknown) true values of birth weight, and we assume that $y_i = f(t_i, x_i) + \epsilon_i$, with $\epsilon_i \stackrel{iid}{\sim} N(0,\sigma^2)$.  Our goal is to sample from the posterior distribution over the regression function $f$ together with the true birth weights $y_i$, given the observed data $(\tilde{y}_i, \gamma_i)$.  

Dropping the $i$ subscript the lighten the notation, we can factorize this joint posterior distribution for $y$ and $f$ as follows:
\begin{align}
p(y, f \mid \gamma, \tilde{y}) &\propto p(y, f) \cdot p(\tilde{y}, \gamma \mid y, f) \nonumber  \\
&\propto p(f) \cdot p(y \mid f) \cdot p(\tilde{y} \mid \gamma, y, f) \cdot p(\gamma \mid y, f) \nonumber \\
&\propto p(f) \cdot p(y \mid f) \cdot p(\tilde{y} \mid \gamma, y)  \label{eqn:babyjointposterior} \, .
\end{align}
The last line follows from two facts: 1) that $\tilde{y}$ is conditionally independent of $f$, given $\gamma$ and $y$; and 2) that the observations are assumed to be coarsened at random, implying that $p(\gamma \mid y, f) \equiv p(\gamma)$ is a constant that can be ignored in sampling from the joint posterior in $(y, f)$.

We now observe that $p(\tilde{y} \mid \gamma, y)$ is a discrete distribution whose PMF and support depend on the coarsening indicator $\gamma$:
$$
\begin{aligned}
p(\tilde{y} \mid \gamma=0, y) &= \mathbb{I}\{ \tilde{y} = y \} \\
p(\tilde{y} \mid \gamma=1, y) &= \mathbb{I}\{ \tilde{y} = \mbox{round}(y, c) \} =   \mathbb{I}\{  \tilde{y}_i - c/2 \leq y_i < \tilde{y}_i + c/2  \} \, ,
\end{aligned}
$$
where $ \mathbb{I}$ is the indicator function, and where the support of the distribution in the second case (where $\gamma = 1)$ is understood to be $\{\tilde{y}: \tilde{y} =  \text{round}(\tilde{y}, c) \}$.  Combining this with the factorized form of (\ref{eqn:babyjointposterior}), and letting $\phi(y \mid m, v)$ denote the density of a normal with mean $m$ and variance $v$, we can write the full joint posterior in $(y,f)$ as
\begin{align}
p(y, f \mid \tilde{y}, \gamma) \propto p(f) &\times \prod_{i: \gamma_i = 0} \phi \left(\tilde{y}_i \mid f(t_i, x_i),\sigma^2 \right) \nonumber \\
&\times \prod_{i: \gamma_i = 1}  \phi \left(y_i \mid f(t_i, x_i),\sigma^2 \right) \mathbb{I}\{  \tilde{y}_i - c/2 \leq y_i < \tilde{y}_i + c/2  \}  \label{eqn:fulljointposterior} \, ,
\end{align}
%
where $p(f)$ is the tsBART prior already described.  Given the regression function $f(t, x)$, the conditional distribution of a true $y_i$ value corresponding to a rounded observation is a truncated normal distribution.  Thus we can easily integrate over uncertainty in $y_i$ in the Bayesian backfitting algorithm: at each MCMC iteration, we draw imputed $y_i$ values for observations where $\gamma_i=1$, using the current MCMC iterates of $f(t_i,x_i)$ and $\sigma$.

\subsection{Projective Smooth BART for monotonicity constraints} \label{sec-meth-b}


We require that model for $f(t,x)$ produces monotone estimates in $t$ for any fixed $x$.  The natural Bayesian strategy for imposing this constraint would be to restrict the prior for $f(t, x)$ to the space of functions that are monotone in $t$, but not necessarily in $x$.  Incorporating this constraint into the prior introduces many computational challenges.  The technical report by \citet{chipman2016b} addresses these challenges for the case of a multiple regression function that is monotone in all its arguments, albeit at the expense of a much more complicated MCMC.  Our model requires monotonicity only in $t$, and incorporates the Gaussian process prior in terminal nodes to induce targeted smoothing in $t$.  \citet{lin2014} show that such a projective prior results in an intractable computational problem even in the ``nice'' the case of a single Gaussian process; this intractability is compounded dramatically in the case of an ensemble of treed Gaussian processes.  

Thus to avoid these computational difficulties, we borrow from the posterior-projection approach in \citet{lin2014}, forming a pseudo-posterior by projecting each posterior sample of $f(t,x)$ to the space of monotone functions in $t$.  That is, if $f(t,x)$ is a posterior draw under our tsBART model, we define
$$
P_f(t, x) = \arg \min_{m \in \mathcal{M}} \int_\mathcal{R} \{ f(t,x) - m(t, x)\} ^2 \ dt \, ,
$$
where $\mathcal{M}$ is the space of functions in $(t,x)$ that monotone non-decreasing in $t$.  Lin et al. project a single Gaussian process and derive a number of properties of the resulting projection operator.   Our regression function is a sum of Gaussian processes in $t$ for any fixed $x$.  Because Gaussian processes are closed under summation, our projection operator inherits these same properties---in particular, if $f(t,x)$ is continuous in $t$ for fixed $x$, then so is $P_f(t, x)$.

Inference is then based on the set of projected posterior draws, $\{P_f(t, x): f \in f^{(1)}, f^{(2)}, \ldots, f^{(S)} \}$, where $S$ is the number of posterior samples and where $f^{(i)}$ is a single draw under the tsBART model without any monotonicity constraints.   It is important to observe that the resulting set of projected draws do not form a Monte Carlo sample from a fully Bayesian posterior.  However, \citet{lin2014} show that, in the case of a single Gaussian process, they correspond to an empirical Bayesian posterior under a mildly data-dependent prior, and that this posterior possesses many appealing theoretical properties.  We note that the study of BART's theoretical properties is an area still in its infancy \citep{castillo2019}. We therefore must leave open the question of whether our prior inherits similar consistency properties as the single-Gaussian-process projection from \citet{lin2014}, but our results show that its empirical performance is excellent.  

We implement the projections using the pooled adjacent violators algorithm (PAVA) \citep{ayer1955, barlow1972, lin2014} which has long been a useful solution in the one-dimensional monotone regression setting \citep{saarela2011}.  PAVA is a computational method for projecting functions to monotonicity; non-monotone adjacent function values are recursively replaced by their mean until the function becomes non-decreasing.   We refer interested readers to \citet{lin2014} for a review of the algorithm.

Finally, we point out that there is another interpretation of our posterior projection approach, in terms of a decision-theoretic outlook on posterior summarization.  Under this interpretation, the unconstrained draws $f$ from the tsBART model form the true Bayesian posterior, but we report a posterior summary $\tilde{f}$ arising from a expected-loss-minimization problem in which we incur infinite loss for non-monotone summaries of the regression function.  While we do not pursue this interpretation in depth, we note that there is a rich tradition of decision-theoretic posterior summarization, e.g. \citet{hahn2015, 7738829, piironen2018projective, woody2019model}; our projection-based approach can be placed in this tradition.  





\begin{centering}\section{Simulations} \label{sec-sims} \end{centering}

We conduct two simulations to illustrate the benefits of incorporating information about coarse observations and monotonicity in the model.  The first simulation is a toy example to provide intuition about the mean squared error inflation incurred when data is coarse.  The second simulation shows performance gains from incorporating monotonicity constraints in appropriate settings. \\

\subsection{Illustration of MSE inflation}

This section uses a toy example to lend intuition to the relationship between degree of coarseness and error in the data, and how they each contribute to mean squared error (MSE) inflation when models do not account for coarseness.  Suppose we have a univariate function of the form $f(x) = \frac{10}{1 + \exp\left(-x \right)} - x$.  We set sample size $N=500$, and we generate univariate $x$'s.  We simulate Gaussian $y \sim N(f(x), \sigma^2)$, and the round the $y$'s to the nearest whole value; $\sigma^2$ is scaled relative to the width of the rounding bins.  

We fit nonparametric regression models using K-Nearest Neighbors and compare the MSE of the fits using the original $y$ versus rounded $\tilde{y}$.  We calculate $MSE$ for the fit using $y$, and $\widetilde{MSE}$ for the fit using $\tilde{y}$.  Define the MSE inflation ratio as
\begin{align*}
	MSE_{IR} =  \widetilde{MSE} / MSE.
\end{align*}

The $x$'s are drawn either uniformly or according to a beta distribution which closely mirrors the distribution of gestational age in our application, and we vary the degree of smoothness in the KNN regression.  We calculate the MSE inflation ratio for 1,000 Monte Carlo draws.

\begin{figure}[ht]
		\centering
			\includegraphics[width=0.6\textwidth]{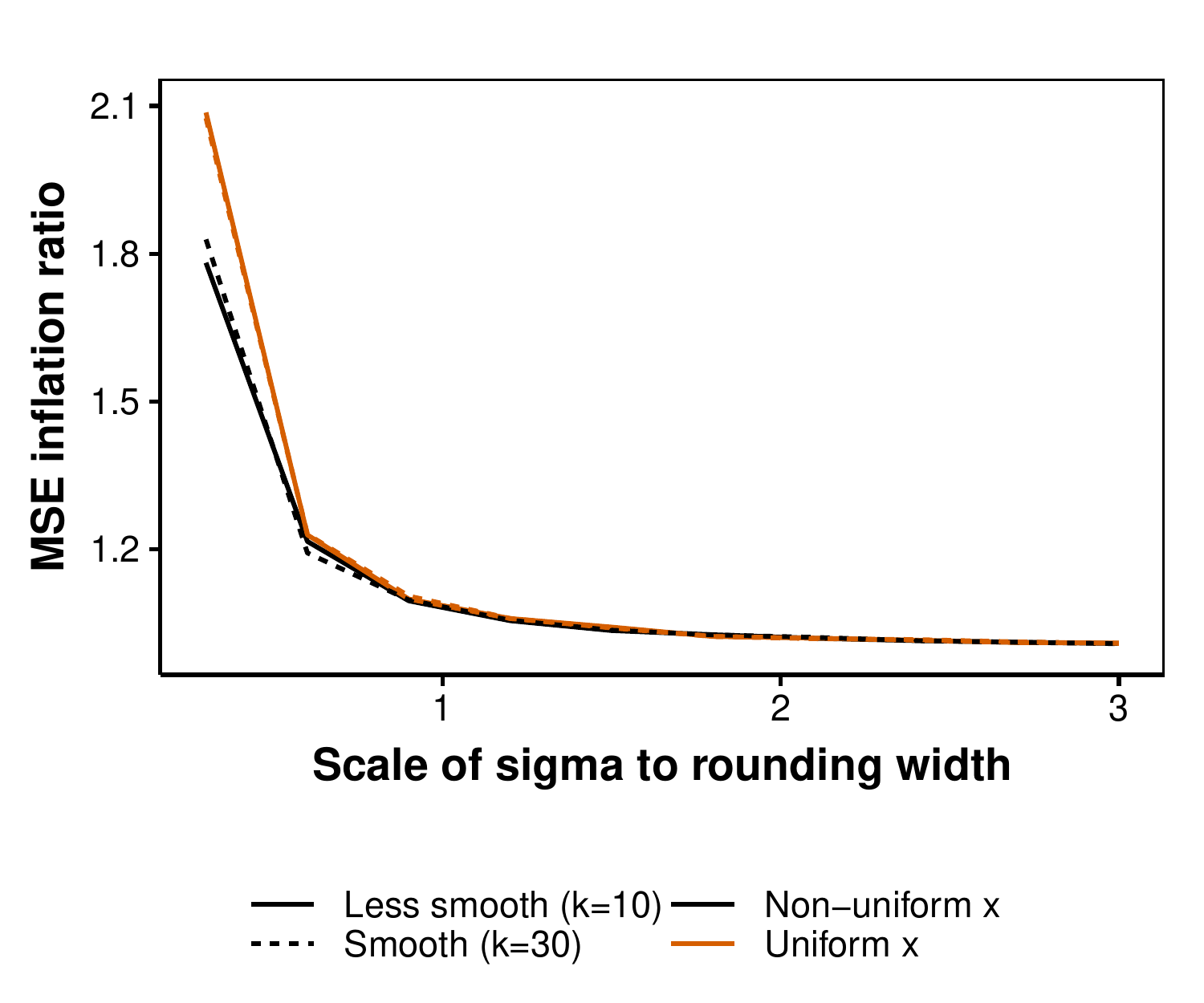}
		\caption{The ratio of MSE when fitting the model using rounded versus non-rounded responses.  On the x-axis, the ratio between rounding bin width and $\sigma$ ranges from 0.3 to 3.  The MSE inflation ratio is calculated for models with and without uniform covariate $x$ and with varying degrees of model smoothness.  MSE inflation depends mainly on the rounding bin width versus $\sigma$ ratio, with uniformity of $x$ and smoothness having smaller impact.}
		\label{fig:mseratio}
	\end{figure}

Figure \ref{fig:mseratio} illustrates our findings related to MSE inflation due to the rounded response.  Primarily, the degree to which MSE inflation is an issue depends on the scale of random error $\sigma^2$ to the width of the rounding bins ($c=1$). We let $\sigma$ vary from 0.3 to 3; when $\sigma$ is smaller than the rounding bin width $c$, MSE inflation is more severe, and when $\sigma \ge c$, the error in the data washes out the impact of rounding on model fit, so MSE inflation is minimal. The uniformity of the $x$'s has a lesser impact, and whether the model is smoothed has even less impact.

\begin{figure}[ht]
		\centering
			\includegraphics[width=0.9\textwidth]{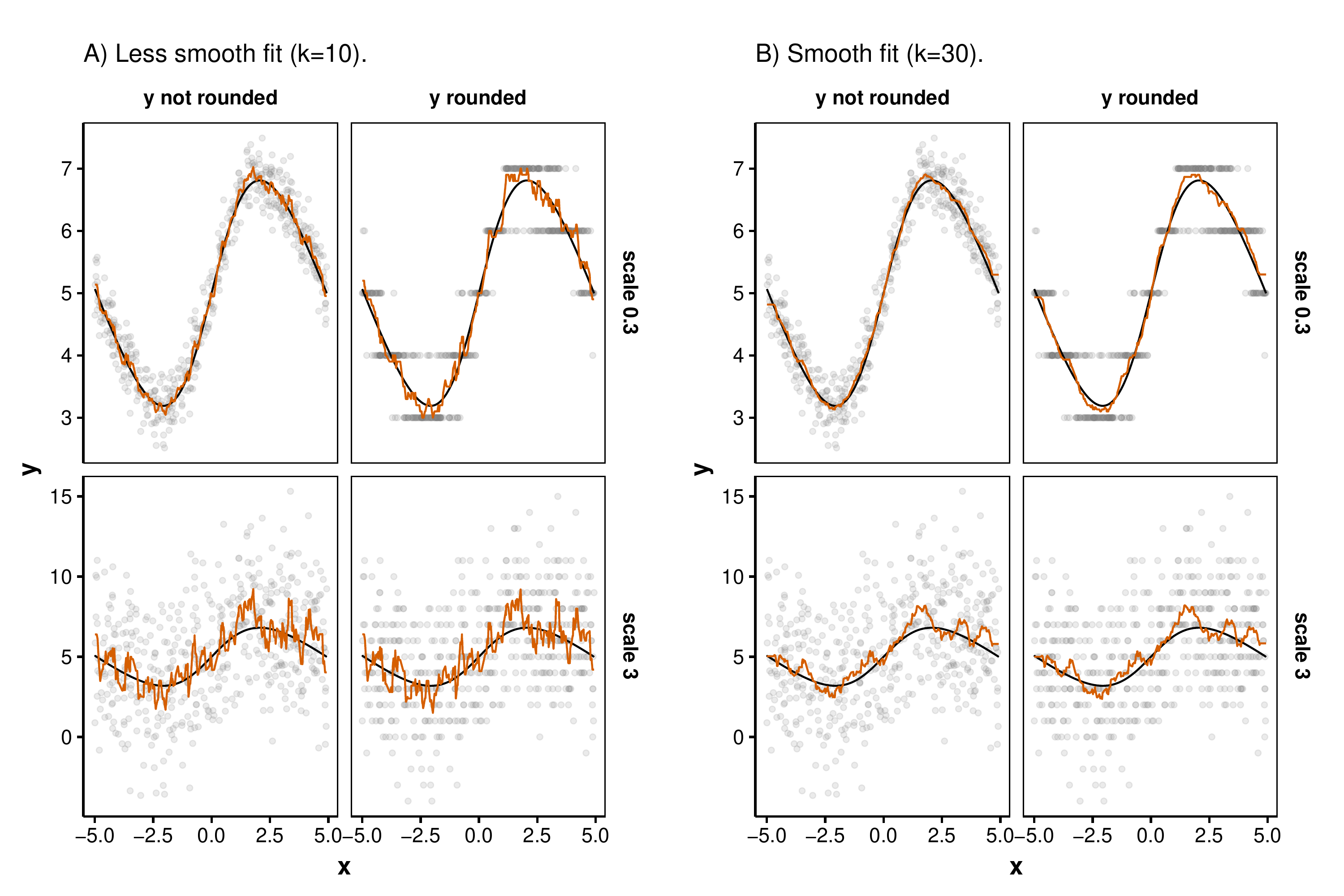}
		\caption{Each panel shows simulated data points $y \sim N(f(x), \sigma^2)$ (grey), the true function $f(x)$ (black line), and the model fit (orange line).  Columns compare results with and without rounded $y$; rows compare scale of $\sigma$ to rounding bin width of 0.3 versus 3.0 (bin width is 1, s.t. $\sigma=.3$ versus $\sigma=3$).  In the top row, $\sigma$ is smaller than the bin width, and the rounded fit is visibly worse than the non-rounded fit; in the bottom row, larger $\sigma$ compared to bin width washes out the rounding impact so that model fits are virtually indistinguishable.  Since the underlying function is smooth, adding smoothing improves fits but does not substantially mitigate rounding issues.}
		\label{fig:toy1}
	\end{figure}

Figure \ref{fig:toy1} lends intuition to support our findings.  Panel A fits  a less-smooth KNN regression ($k=10$ neighbors considered), and Panel B sets $k=30$ for a smoother fit.  Each panel shows simulated data points $y \sim N(f(x), \sigma^2)$ (grey), the true function $f(x)$ (black line), and the model fit (orange line).  Columns compare results with and without rounded $y$; rows compare scale of $\sigma$ to rounding bin width of 0.3 versus 3.0 (bin width is 1, s.t. $\sigma=.3$ versus $\sigma=3$).  In the top row, $\sigma$ is smaller than the bin width, and the rounded fit is visibly worse than the non-rounded fit; in the bottom row, larger $\sigma$ compared to bin width washes out the rounding impact so that model fits are virtually indistinguishable.  Since the underlying function is smooth, adding smoothing improves both fits commensurately but does not substantially mitigate rounding issues.

\subsection{Incorporating monotonicity}
This section presents the results of a simulation that illustrates the advantage of incorporating monotonicity constraints in scenarios where the underlying function is monotonic.  We simulated datasets which are monotone in the target covariate $t_i \in \left\{1,2,\hdots,10 \right\}$, with two covariates, $x_{1i} \stackrel{iid}{\sim} \text{U}(0.5, 1.5)$ and $x_{2i} \stackrel{iid}{\sim} \text{U}(0,1)$.  We consider three choices for $f(t, x_1, x_2)$:
\begin{itemize}
	\item Arctan: $f(t,x_1,x_2) = 2 + 0.5x_2 + \left(1 + \exp\left(-x_1 (t-5) \right) \right)^{-1}$
	\item Linear: $f(t,x_1,x_2) = 2 + \text{atan}(t) + 0.25x_1 (\frac{t}{5}) - 0.5x_2$
	\item Sigmoid: $f(t,x_1,x_2) = 2 + \frac{t}{8x_1} + 0.5x_2$
\end{itemize}

\begin{figure}[t]
		\centering
			\includegraphics[width=0.9\textwidth]{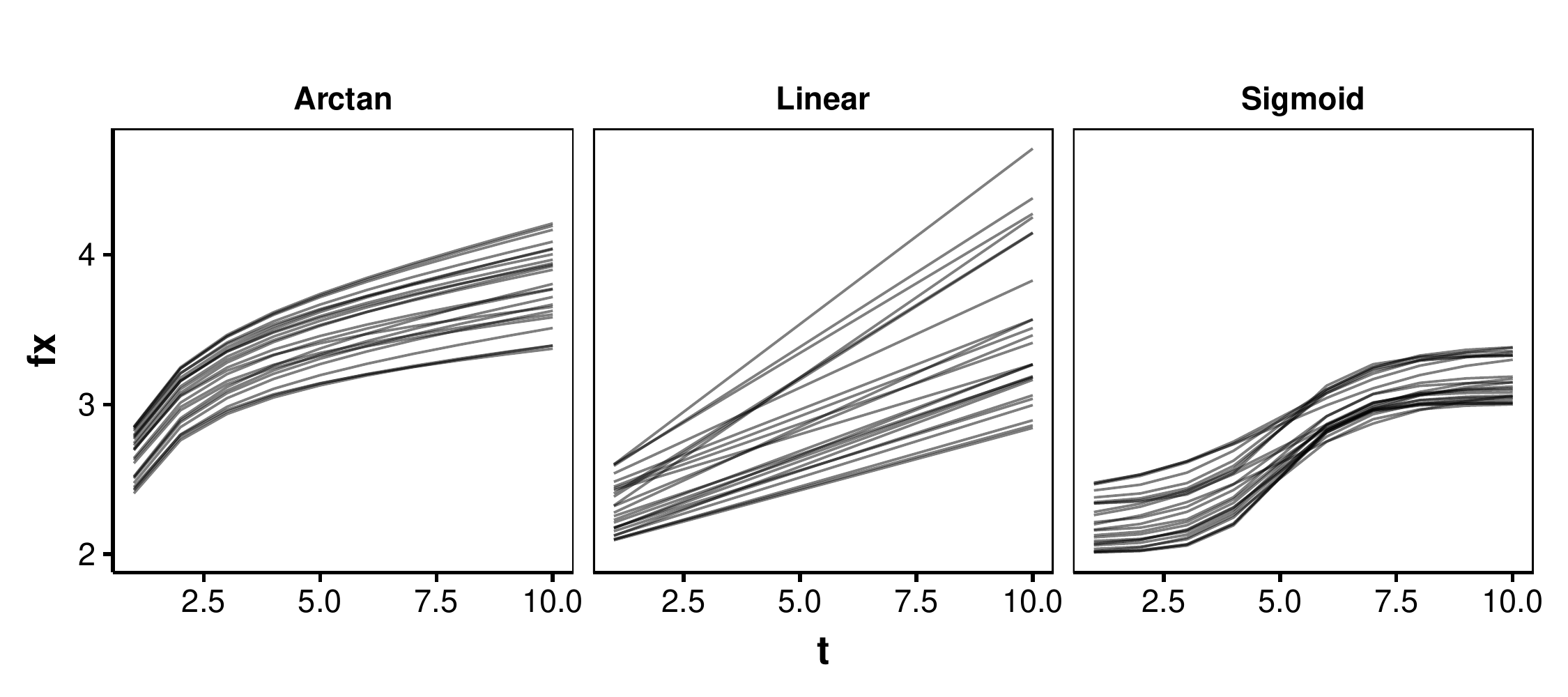}
		\caption{Scenarios for the simulation.  Each panel shows a function choice, with lines representing $(x_1,x_2)$ draws over $t$.}
		\label{fig:monosim-figure-01_scenarios}
	\end{figure}

These choices are shown in Figure \ref{fig:monosim-figure-01_scenarios}, where we plot draws of $(x_1,x_2)$ over target covariate $t$. The functions increase monotonically over $t$ in varying ways, but are not monotonic increasing over remaining covariates.

We sample 50 datasets of 1,000 observations each, calculate $f(t,x_1,x_2)$ for each scenario, and generate observations using standard Gaussian error.  Each scenario is fitted using an 80/20 test-train split.  We fit each scenario for each dataset twice; once using tsBART out of the box, and again using tsBART with a monotone increasing constraint.  We calculate the mean out-of-sample RMSE across replicates for each scenario.

\begin{table}[ht]
	\caption{\label{tab:monosim-results}
	Results from monotonicity simulation.  Compares the out-of-sample MSE, averaged over 50 replicates, for fitting the tsBART model with default settings versus constrained to monotone increasing estimates over the target covariate.  Enforcing 	monotonicity when present provides small performance gains}
\centering
\begin{tabular}{rlrrr}
  \hline
 & Scenario & MSE Default & MSE Monotone & Percent MSE Reduction \\ 
  \hline
1 & Arctan & 1.03 & 1.02 & 1.20 \\ 
  2 & Linear & 1.03 & 1.02 & 1.07 \\ 
  3 & Sigmoid & 1.04 & 1.03 & 0.75 \\ 
   \hline
\end{tabular}
	
\end{table}

Results are shown in Table \ref{tab:monosim-results}, where we find that imposing the monotonicity constraint results in modest but beneficial reductions in MSE for all scenarios.  These small gains are to be expected, given that the BART framework already provides excellent predictive power.  The most persuasive reason for enforcing monotonicity is for reasons of interpretability in scenarios where deviations from monotonicity are clinically implausible or impossible, and where patients and/or clinicians are likely to over-interpret small wiggles in a nonparametrically estimated regression function.  We view small performance gains as an additional ancillary benefit.

\begin{centering}\section{Results for Modeling Birth Weight} \label{sec-results} \end{centering}


We now return to our motivating application, investigating the impact of pre-eclampsia on birth weight in sub-Saharan Africa.  The target covariate for monotonicity and smoothing is gestational age of delivery in weeks: $t_i \in \left\{28,\ldots,42 \right\}$.  Response $\tilde{y}_i$ is the birth weight (kg), and $\gamma_i$ indicates whether $\tilde{y}_i$ may be rounded to the nearest 0.1 kg.
\begin{align*}
	\gamma_i = \begin{cases}
		1, \quad \text{if } \tilde{y}_i \text{ mod } 0.1 = 0 \\
		0, \quad \text{otherwise.}
	\end{cases}
\end{align*}
Let $x_i$ be the vector of maternal-fetal covariates for each patient, including: presence of maternal pre-eclampsia, maternal age, sex of the infant, maternal job status, Ganda ethnicity of the mother, maternal HIV status, and parity.

A key feature of the application is that, for a fixed level of the covariates $x$, expected birth weight should not decrease with gestational age.  (Note that this is not necessarily true when comparing pregnancies with different values of $x$.)  We use the projection approach previously described to enforce this constraint.   We also impute ``de-coarsened'' $y_i$ values where $\gamma_i=1$ as described in Section \ref{sec-meth-c}.  We then calculate posterior mean predicted birth weight estimates for each patient, after projecting posterior draws for $f(t,x)$ to the space of functions monotone in $t$.  

\begin{figure}[t]
		\centering
			\includegraphics[width=0.9\textwidth]{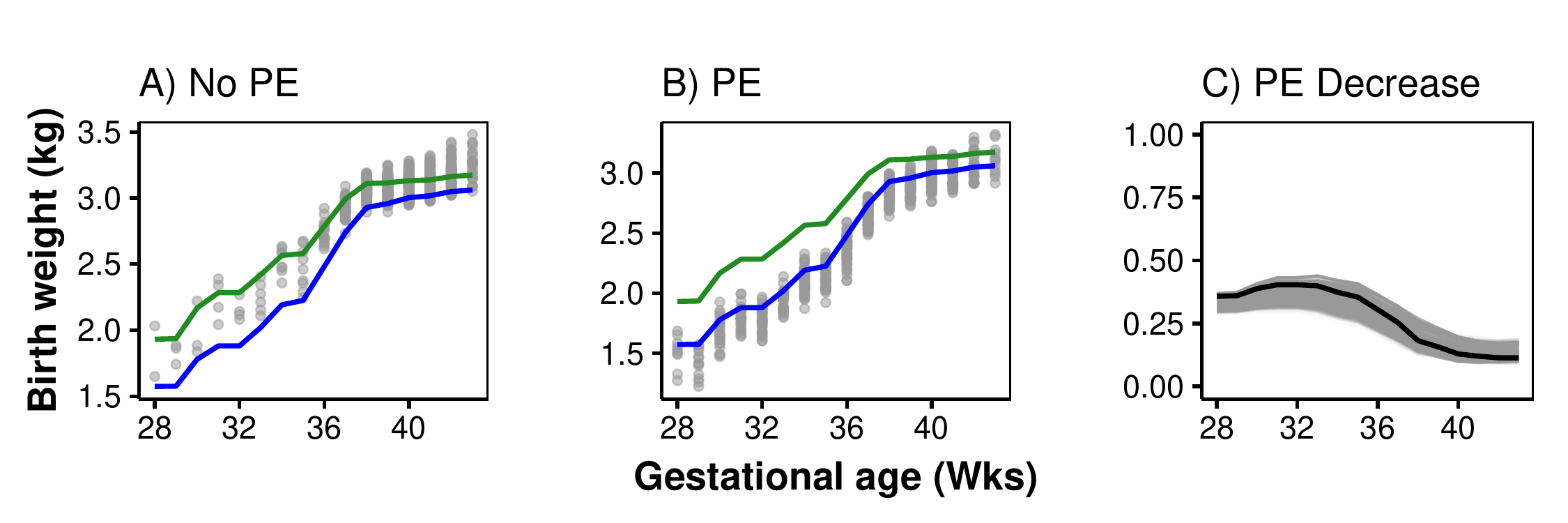}
		\caption{Posterior mean predicted birth weights and differences associated with PE, across gestation.  Panels A and B give pointwise posterior mean birth weight estimates for individuals; solid lines represent estimated paths acros gestation for a `centroid' patient with the most common features of the dataset, with (blue) and without (green) pre-eclampsia.  Panel C shows estimated decrease in birth weight associated with pre-eclampsia for all patients over gestation (grey), and for the centroid patient (black). }
		\label{fig:mulag-insamp}
	\end{figure}

Our goal is to characterize the relationship between birth weight and pre-eclampsia across gestation in urban Uganda, and provide clinicians with insight into the comparative amount of influence pre-eclampsia has on birth weight compared to other maternal-fetal covariates.   Figure \ref{fig:mulag-insamp} plots posterior mean birth weight estimates for each patient, with normotensive patients in Panel A, and pre-eclamptic patients in Panel B.  Both panels show the posterior mean birth weight estimates for a `typical' patient under normotensive (green) and pre-eclamptic (blue) conditions.  Here we define `typical' as the centroid of covariate space, i.e.~a patient whose maternal covariate values are the means (and modes, for categorial features) represented in the data.  Panel C shows the estimated decrease in birthweight across gestation associated with preeclampsia for the centroid patient (black line); we see a clear decrease in birth weight associated with pre-eclampsia across all gestations, with a slightly more drastic difference earlier in pregnancy.  Specifically, if we partition maternal-fetal covariates $x$ as $x = (\text{pec}, v)$, where pec is an indicator for preeclampsia and $v$ is all other covariates, the black line in Panel C shows the quantity
$$
\Delta(t, v_0) = \hat{f}(t, \text{pec}=1, v_0) - \hat{f}(t, \text{pec}=0, v_0)
$$
where $v_0$ is the centroid of covariate space.  We also calculate the quantity $\Delta(t, v_i)$ for each observed combination of covariates $v_i$, and the grey strip in Panel C shows the range of $\Delta(t, v_i)$ across the entire data set for each gestational age.  

\begin{figure}[t]
		\centering
			\includegraphics[width=0.9\textwidth]{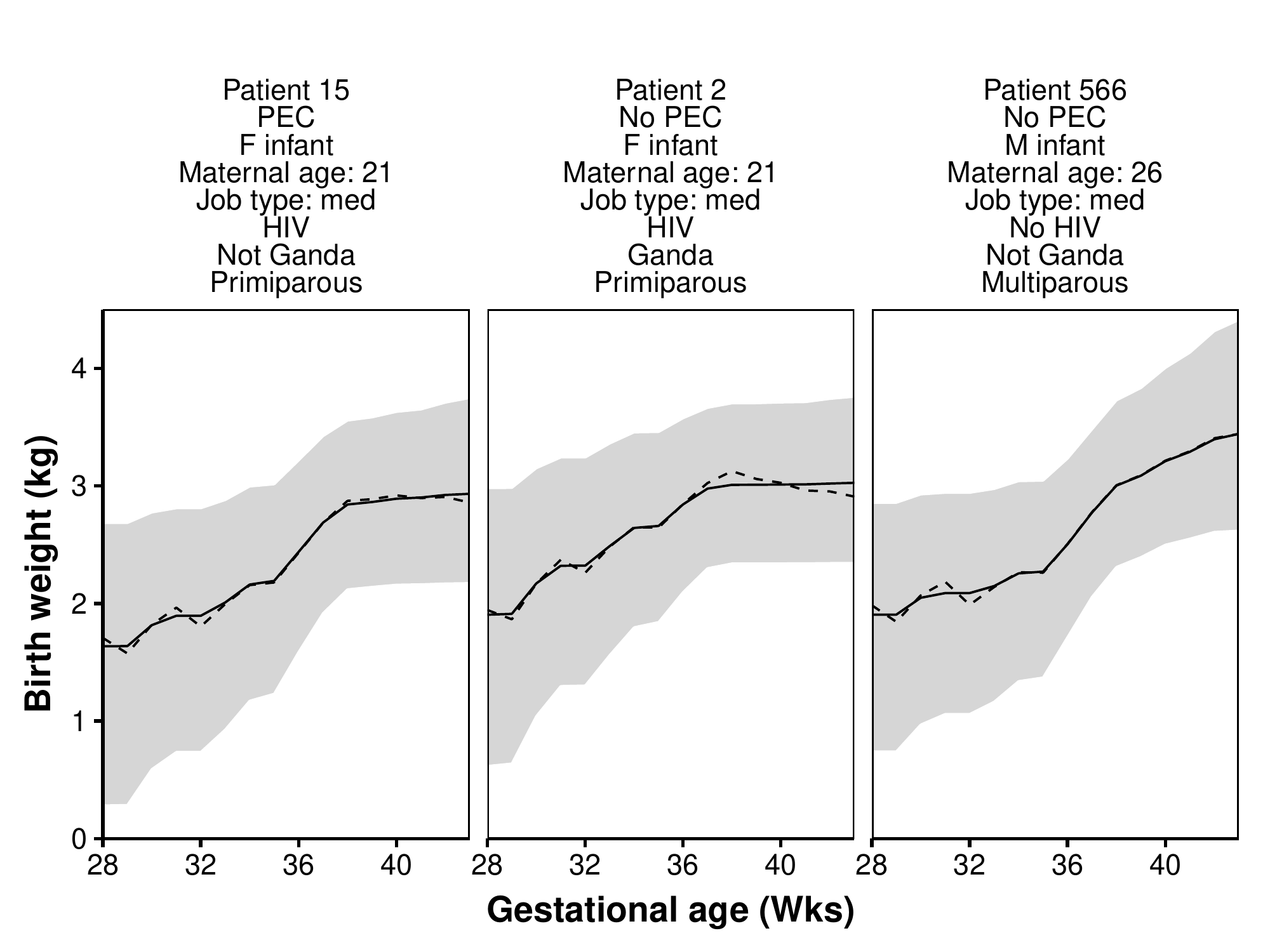}
		\caption{Expected birth weights over gestation for three randomly chosen combinations of $x$, representing three hypothetical patients.  Lines represent posterior mean birth weight curves over gestation, with (solid) and without (dashed) monotonicity constraints.  The 95\% posterior prediction intervals for the monotone case are shaded. We see that inducing monotonicity is not substantively altering the estimates of birth weight, but is smoothing small wiggles in the growth curves due to variance, making estimated birth weight trajectories more interpretable for clinicians without sacrificing accuracy.}
		\label{fig:mulag-spaghetti}
	\end{figure}

Figure \ref{fig:mulag-spaghetti} illustrates the subtle yet critical impact of monotonicity constraints in this setting.  Each panel is a randomly chosen combination of $x$ values representing some patient.  Lines represent posterior mean birth weight curves over gestation, with (solid) and without (dashed) monotonicity constraints.  The 95\% posterior prediction intervals for the monotone case are shaded. We see that inducing monotonicity is not substantively altering the estimates of birth weight, but is smoothing small wiggles in the growth curves due to variance, making estimated birth weight trajectories more interpretable for clinicians without sacrificing accuracy.

To further investigate the relationship of pre-eclampsia to birth weight, we use a ``fit-the-fit'' approach to explore which covariates (and subgroups of covariates) seem to be driving most of the variation in expected birth weight.  A typical approach is to fit a CART model \citep{chipman1998} to the BART output, with posterior means $\hat{f}(t_i, x_i)$ for each observation as the response, and maternal-fetal covariates as predictors.  Because pre-eclampsia is associated with early delivery \citep{nakimuli2019}, we deconfound this by fitting the CART model to $r_i$, the posterior mean log percent differences conditional on time.
\begin{align*}
	r_i = \log \left[ 
		\hat{y}(t_i,x_i) \ \bar{y}_{t=t_i}
	\right]
\end{align*}
where $\bar{y}_{t=t_i}$ is the mean of observed birth weight values values for patients who also delivered at gestational age $t_i$, and $\hat{y}(t_i,x_i)$ is the posterior mean birth weight for observation $i$.  This approach finds subgroups of covariates associated with being born small given delivery at a particular age, as opposed to modeling deviations in birth weight across all gestations, which would capture covariates like pre-eclampsia which are strongly associated with gestational age, but not necessarily with being small or large for a given gestational age.

\begin{figure}[t]
		\centering
			\includegraphics[width=0.9\textwidth]{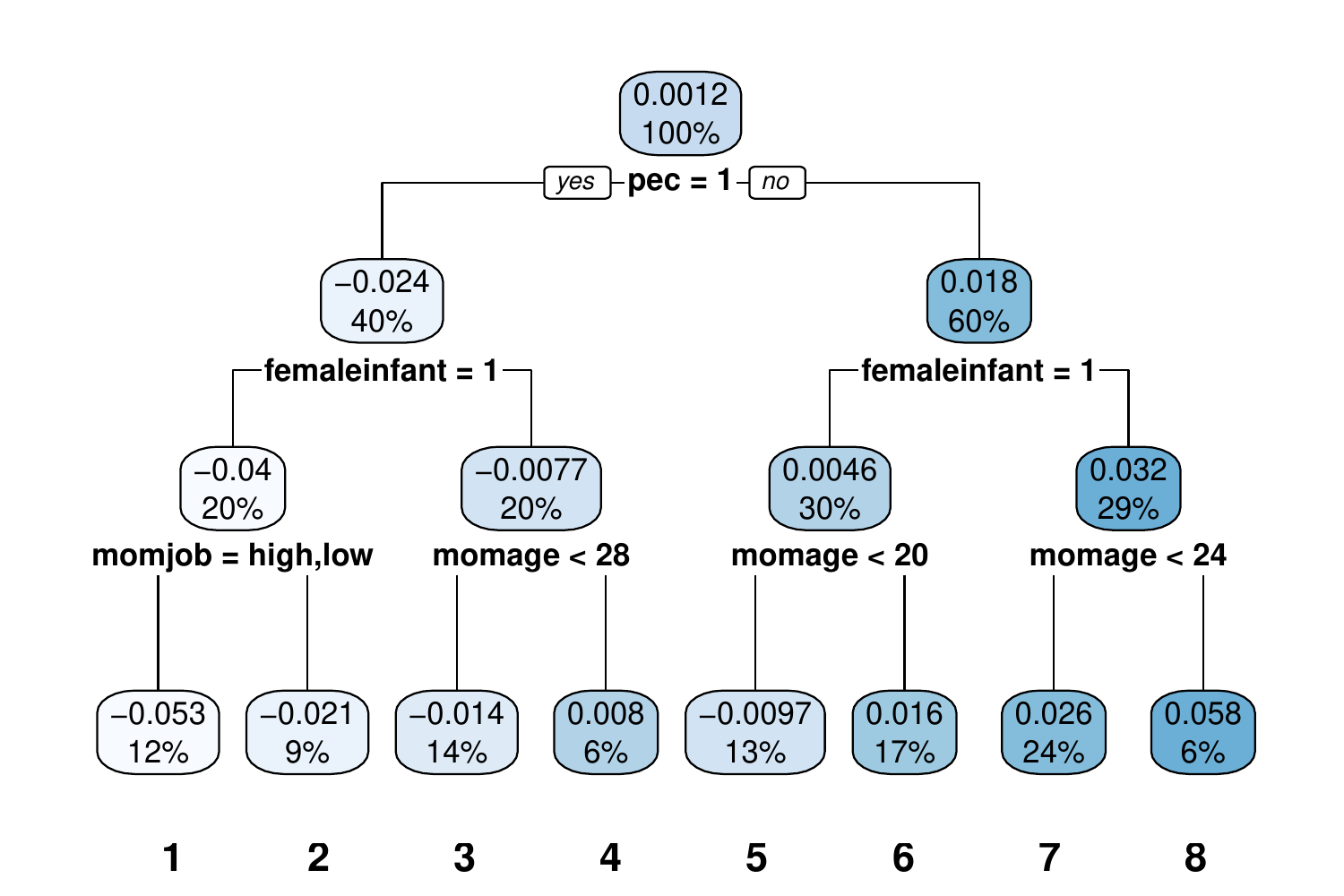}
		\caption{The tree fit from the CART model, fitting posterior mean log percent difference in birth weight conditional on gestational age.  Pre-eclampsia is the primary split, with pre-eclampsia cases associated with lower birth weights; followed by infant sex, and then maternal age, with female babies and younger mothers linked to lower birth weight.  Pre-eclampsia is the dominant predictor in being underweight compared to other deliveries at the same point in gestation.}
		\label{fig:mulag-cart}
	\end{figure}
	
Figure \ref{fig:mulag-cart} gives the tree fit from the CART model.  Each tree node in the figure contains the posterior mean log percent difference and the percent of observations falling in that node.  Nodes are numbered for reference from left to right.  Pre-eclampsia, the first tree split, is the most important subgroup, with pre-eclamptic pregnancies resulting in lower birth weights.  For pregnancies both with and without pre-eclampsia, sex of the infant is the next most impactful covariate -- female infants tend to be smaller than males.  Within both the male and female cohorts in each branch of the tree, increase in maternal age is indicative of higher birth weights, a finding which confirms previous studies \citep{bakker2011}.

\begin{figure}[t]
		\centering
			\includegraphics[width=0.9\textwidth]{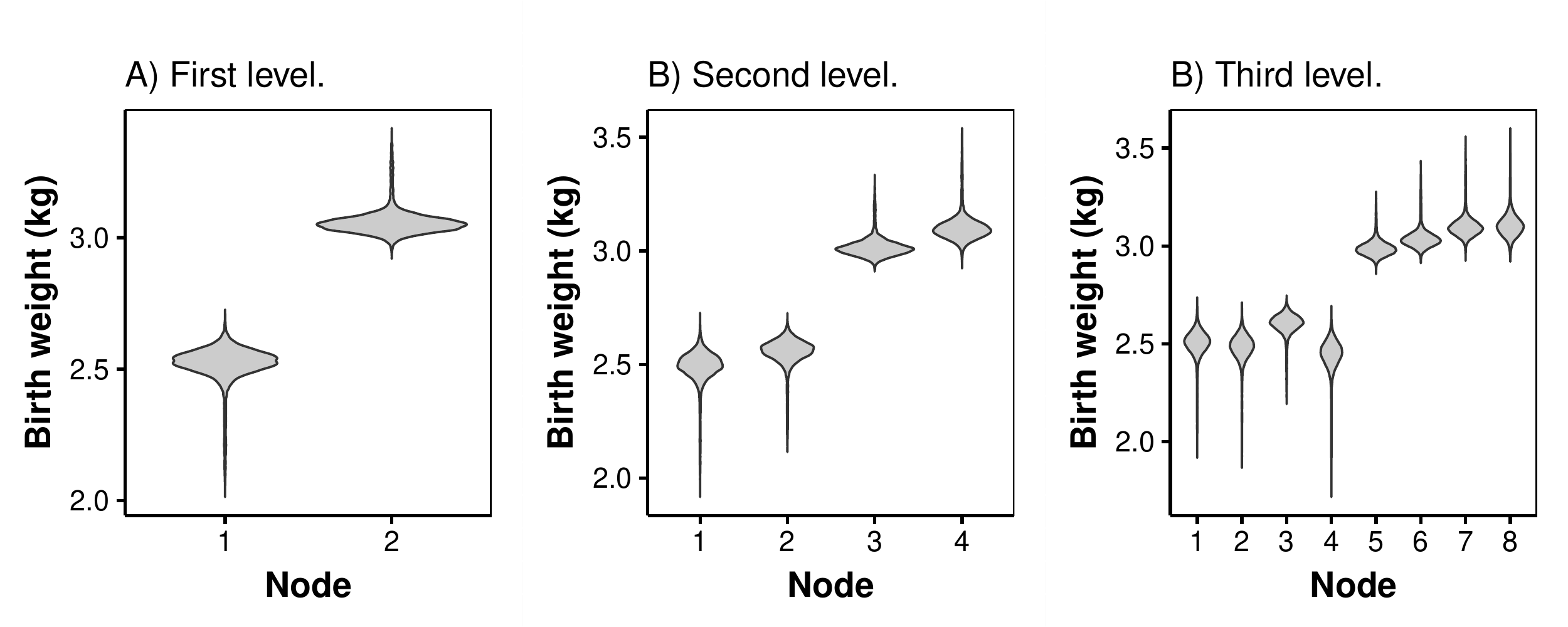}
		\caption{Posterior birth weight densities for each level of CART tree splits.  There is clear separation between the posteriors for patients with and without pre-eclampsia, indicating its dominant role in predicting birth weight.  Infant sex also plays a role; other covariates appear to be far less important in predicting birth weight.}
		\label{fig:mulag-cart-posteriors}
	\end{figure}

The CART fit tree is useful for identifying subgroups, but tree nodes capture point estimates; without capturing uncertainty it is difficult to discern how different tree splits are from each other, and to what extent each covariate in the tree is driving birth weight.  We query posterior birth weight draws for nodes at each level of the CART tree, and plot the posterior densities in Figure \ref{fig:mulag-cart-posteriors}.  Less overlap in posterior densities indicates meaninful splits.  Concentration of the posterior densities indicates more precisely isolated subgroups as a result of the tree split, and dispersion indicates greater heterogeneity within nodes at a given tree level. We refer to tree nodes at each level in numeric order from left to right.

In Figure \ref{fig:mulag-cart-posteriors}, Panel A shows posterior densities for patients with pre-eclampsia (node 1) versus without pre-eclampsia (node 2).  There is clear separation between these two subgroups, indicating that pre-eclampsia is having a marked impact on birth weight.  Panel B gives posterior densities for each infant sex within pre-eclampsia groups, where nodes one and three are female infants, and two and four are male.  There is far more overlap between these groups than the pre-eclampsia group, and we notice that infant sex plays a slightly more important role for non-pre-eclampsia patients, while there is more overlap in the pre-eclampsia group between sexes.  Panel C shows posterior densities for the bottom level of the tree, with one posterior density per terminal node.  These posteriors are clustered closely, with a large amount of overlap within each pre-eclampsia group.

To summarize these findings, pre-eclampsia is the dominant covariate in predicting birth weight.   Patients with pre-eclampsia deliver babies at lower birth weights at any given gestational age than normotensive patients at the same gestational age.  Pre-eclampsia is associated with a slightly larger decrease in birth weight at earlier gestational ages.  Within cohorts of normotensive and pre-eclamptic patients, infant sex is the next best predictor of birth weight, with female babies being lighter in both cohorts. \\

\begin{centering}\section{Discussion} \label{sec-discussion} \end{centering}


Analysis of data from the Mulago National Referral Hospital has demonstrated that maternal pre-eclampsia is the dominant influence on birth weight in the urban Uganda setting.  Our work represents substantial advancement on previous analyses \citep{nakimuli2019}, as we consider smooth monotone function estimation and accounting for uncertainty due to data coarsening via rounded birth weight estimates, as well as in capturing heterogeneity of growth curves by patient.  Our Bayesian posterior analysis using log-deviation of estimated birth weight conditional on gestation allows for delineation of maternal-fetal covariates while carefully deconfounding the association between pre-eclampsia and early delivery.  The posterior analysis demonstrates that pre-eclampsia is the dominant influence on decreased birth weight given gestational age at delivery.  We find the amount of decrease in birth weight associated with pre-eclampsia to be slightly larger at earlier gestations.  We also find evidence that for both pre-eclamptic and normotensive patients, infant sex plays a role in birth weight, with female babies being smaller on average.  We find that other elements of the maternal-fetal dyad have much smaller impact on birth weight than pre-eclampsia.

The baseline risk of neonatal death in LMIC is high, at 30 per 1,000 in sub-Saharan Africa \citep{who2014}, and is raised considerably by being either small for gestation or pre-term \citep{katz2013}.  A pooled analysis by \citet{katz2013} shows that across a range of LMICs, the combined effect of being born both small and early increases risk of neonatal mortality over 15-fold above baseline.  Our results provide insight into a common cause of being delivered both small and early, which may be useful in developing heuristics which can guide clinical practice and be easily implemented in a busy urban sub-Sahara Africa hospital setting.  Proactive identification of pregnancies most likely to result in low birth weight allows clinicans to initiate appropriate interventions to reduce incidence of low birth weight and so decrease perinatal mortality. \\

\begin{centering}\section{Supplemental Materials} \label{sec-supp} \end{centering}


The tsBART R package, including implementation of all methods described in this work, can be round at \url{https://github.com/jestarling/tsbart}.  Code to reproduce tables and figures is available at \url{https://github.com/jestarling/mulago-analysis}.\\
	
\begin{centering}\section*{REFERENCES}\end{centering}
\singlespace
\begin{small}\
	\bibliographystyle{abbrvnat}
	\bibliography{mulago-paper}

\end{small}
		

\end{document}